\pdfoutput=1
\documentclass[useAMS,usenatbib,usedcolumn]{mn2e}
\usepackage{rotating}
\usepackage{lscape}
\usepackage{latexsym}
\usepackage{graphicx}
\usepackage{sidecap}
\usepackage{acronym}
\usepackage{textcomp}
\usepackage{bbding}
\usepackage{epsf}
\usepackage{amsmath,amssymb}
\usepackage{float}
\usepackage{multirow}
\usepackage{flafter}
\usepackage{placeins}
\usepackage{afterpage}
\usepackage{preview}
\usepackage{here}
\usepackage{float}
\usepackage{sidecap}
\usepackage{setspace}
\usepackage{hyperref}
\usepackage[bottom]{footmisc}
\usepackage{lipsum}
\usepackage{lscape}
\usepackage{fixltx2e}
\hypersetup{pdfauthor={M.Stupar},
pdftitle={ Confirmation of G6.31+0.54 as a part of a Galactic supernova remnant},
pdfkeywords={supernova remnants, Galaxy: structure, ISM: structure},
bookmarksnumbered=true}
\newcommand{\D}{$^\circ$}
\newcommand{\NII}{[N\,{\sc ii}]}

\newcommand{\SII}{[S\,{\sc ii}]}
\newcommand{\HA}{H{$\alpha$}}
\newcommand{\OIII}{[O\,{\sc iii}]}

\newcommand{\OII}{[O\,{\sc ii}]}

\newcommand{\HB}{H{$\beta$}}

\def\p0{\phantom{0}}

\def\k{km s$^{-1}$}
\def\ks{km s$^{-1}$}

\def\cm3{cm$^{-3}$}

\def\12{$^{12}$CO}
\def\13co{$^{13}$CO}

\def\HII{H{\sc ii}}

\def\la{{\hbox{$\lesssim$}}}

\title[SNR optical confirmation of G6.31+0.54]
{Confirmation of G6.31+0.54 as a part of a Galactic supernova remnant}
\author [Stupar,  Parker \& Frew]
{M.~Stupar,$^{1,2}$ \& Q. A.~Parker,$^{1,2}$ and D.J.~Frew$^{1,2}$\\
\\
$^{1}$Department of Physics, University of Hong Kong, CYM Physics Building, Hong Kong\\
$^{2}$The Laboratory for Space Research, University of Hong Kong, Hong Kong\\}
 \date{Accepted 201;
      Received 2017;
     in original form 2017}
 \volume{{\rm in press}}
\begin{document}

\maketitle
\begin{abstract}  
A combination of archival multi-frequency radio observations with narrow-band H$\alpha$ optical imagery and new confirmatory optical spectroscopy 
have shown that candidate supernova remnant G6.31+0.54 can now be confirmed as part of a Galactic supernova remnant (SNR). It has non-thermal 
emission, an optical emission line spectrum displaying shock excitation and standard SNR line ratios, fine filamentary structures in \HA\ typical of optical 
remnants and closely overlapping radio and optical footprints. An X-ray ROSAT source \mbox{1RXS J175752.1-231105} was also found that 
matches the radio and optical emission though a definite association is not proven. 
Nevertheless, taken together, all these observed properties point to a clear SNR identification for 
this source. We provide a rough estimate for the kinematic distance to G6.31+0.54 of $\sim$4.5~kpc. The detected optical filaments are some $\sim$~10~arcminutes  in extent (or about 13~pc at the assumed distance). However, as only a partial arcuate structure of the SNR can be seen  (and not a full shell) the full angular extent of the SNR is unclear. Hence the physical extent of the observed partial shell is also difficult to estimate. If we assume an approximately circular shell then a conservative fit to the optical arc shaped 
filaments gives an angular diameter of $\sim$20~arcminutes corresponding to a physical diameter of  $\sim$26pc that shows this to be an evolved remnant.

 \end{abstract}


\section{Introduction}

Only \mbox{$\sim280$} Galactic supernova remnants are currently known \citep{gree14} while studies of Galactic SN explosion rates 
(1.9/per century, e.g. \citet{capp99}) combined with pulsar birth rates (1 expected every 30 to 250 years e.g. \citealt{grrim04}) and their 
average 100,000 year lifetimes, indicate there could be $\sim$1000 SNR currently present in the Galaxy \citep{cas98}. Hence, a significant 
fraction either remains to be uncovered or our present understanding of SNR birth rates are out by a factor of 2-3. Either way the detection and 
confirmation on new Galactic supernova remnants is an important task. 

This is the latest  in series of follow-up papers (e.g. \citealt{stu07}; \citealt{stu08}; \citealt{stu09}; \citealt{stu11} and \citealt{sab13}) 
that have uncovered a significant new sample of optical detections for Galactic SNRs based on narrow-band \HA\ imagery from the SuperCOSMOS 
UKST/AAO \HA\ survey of the Southern Galactic plane (SHS hereafter: \citealt{parker05}; \citealt{Frew14}) or from the IPHAS \HA\ survey of the 
Northern plane  \citep{drew14}.  Optical counterparts to radio detected SNRs have had more prominence recently due to the 
advent of such wide-field, high sensitivity and fine resolution \HA\ surveys of the Galactic plane where they also 
form an SNR discovery medium in their own right (e.g. \citealt{stu08}). The detection of SNRs in the optical, mostly in \HA\, is usually in the form of fine 
filamentary structures of different  shape and size. Often they have the same overall morphological form as seen in the radio, infrared or X-ray images with 
overlapping complete or partial common footprints. However, it is also the case that filaments or diffuse  \HA\ emission  can be seen 
separately, without any obvious direct correlation with other wavebands  (e.g. see certain  objects in \citealt{stu08} or \citealt{Fes10}). 

\citet{bou04, bou05, bou08, bou09, Fes10}  have also published first optical detections and subsequent spectroscopic confirmation for several additional  SNRs 
based on CCD imagery. Multi-wavelength SNR investigations are also needed and are essential for understanding shock behaviour in the ISM, examining the 
correlation of shocked plasma with dust, non-thermal and thermal processes, the behaviour of radiative and non-radiative shocks, elemental abundances and 
heavy element ISM enrichment. The probable connection of Balmer shocks and X-ray/gamma ray emissions in the case of cosmic ray acceleration is also an important multi-frequency task.  

This paper presents a new multi-wavelength investigation of SNR candidate G6.31+0.54  in a similar fashion to  \cite{stu07} and \cite{stu11}.  It 
was first noted a possible SNR by \cite{bro06} as one of their ``class III" candidates with positional coincidence with non-thermal emission 
but that required further deeper and more detailed follow-up for confirmation. It can also just be discerned in the 20~cm Multi-Array Galactic Plane Imaging Survey (MAGPIS) - \citealt{helf06} as source 6.3333+0.5500. Here we present multi-frequency radio observations from 
available radio surveys together with the first reported \HA\ optical detection.  Integral field unit (IFU) 3-D spectroscopy at various positions across the optical 
filaments was undertaken and the integrated 1-D spectra are also presented.  We discuss the probable association of G6.31+0.54 with a ROSAT X-ray source 
recognized within the region encompassed the by radio and optical data of the SNR.

\section{Radio observations}
\subsection{First radio detection of G6.31+0.54}

This possible SNR was first noted by \citet{bro06} among 35 new SNRs and candidates reported using radio observations made by the VLA in 
the \mbox{B,~C and D} configurations at 90~cm. This was combined with other archival data from different surveys at 20~cm (e.g from the Southern 
Galactic Plane Survey (SGPS), observed with the Australia Telescope Compact Array \citep{mcC05}  and at 11~cm  from the  4.3~arcminute  higher 
resolution single-dish Bonn survey data (\citealt{reich90}). See also further details in \cite{bro06} where they established three main criteria for 
identification of an SNR at 90~cm: i) the object must be resolved, showing a whole or at least a reasonable and identifiable fraction 
of a typical SNR shell; ii)  the radio continuum spectral index $S_{\nu}$~$\propto$~$\nu^{\alpha}$ must be negative\footnote{where frequency is $\nu$, radiative flux density is $S_{\nu}$ and the spectral index is $\alpha$} , presenting non-thermal 
(synchrotron) emission and iii) the source must be distinct from mid-infrared emission at 8~$\mu$m \citep[compared with MSX 8~$\mu$m sensitivity, 
e.g.][]{price01}.\footnote{This is due to fact that \HII~regions and wind-blown bubbles are similar in morphological structure to SNRs, with flat radio 
spectra but surrounded with bright 8~$\mu$m emission.} 

\citet{bro06} divided the new SNR candidates into three classes (I,~II or III) according to confidence they are real SNRs. G6.31+0.54 
satisfied two criteria:  the spectral index across three wavelengths  of 11, 20 and 90~cm was indeed negative (\mbox{ $\alpha_{(90/20)}$=~--0.5} and  \mbox{$
\alpha_{(90/11)}$=--0.8}) and typical for SNRs. No association was found between the radio and mid infrared emission from the MSX 8~$\mu$m band 
\citep[for feasible  optical/radio/MSX 8~$\mu$m relationship see also][]{stu11}. The third condition was not 
completely satisfied so G6.31+0.54 was assigned class III for sources: ``coincident with non-thermal emission but very faint, 
very confused, or do not exhibit a typical shell-type SNR morphology". Follow-up of this candidate was considered ``essential".  \citet{gree14} does not 
include this object in his Galactic SNR catalogue.

\subsection{G6.31+0.54 as seen at 6~cm (PMN) and 20~cm (NVSS and MAGPIS)}
\begin{figure}
\center
%Fig.1
\includegraphics[width=230pt,height=180pt]{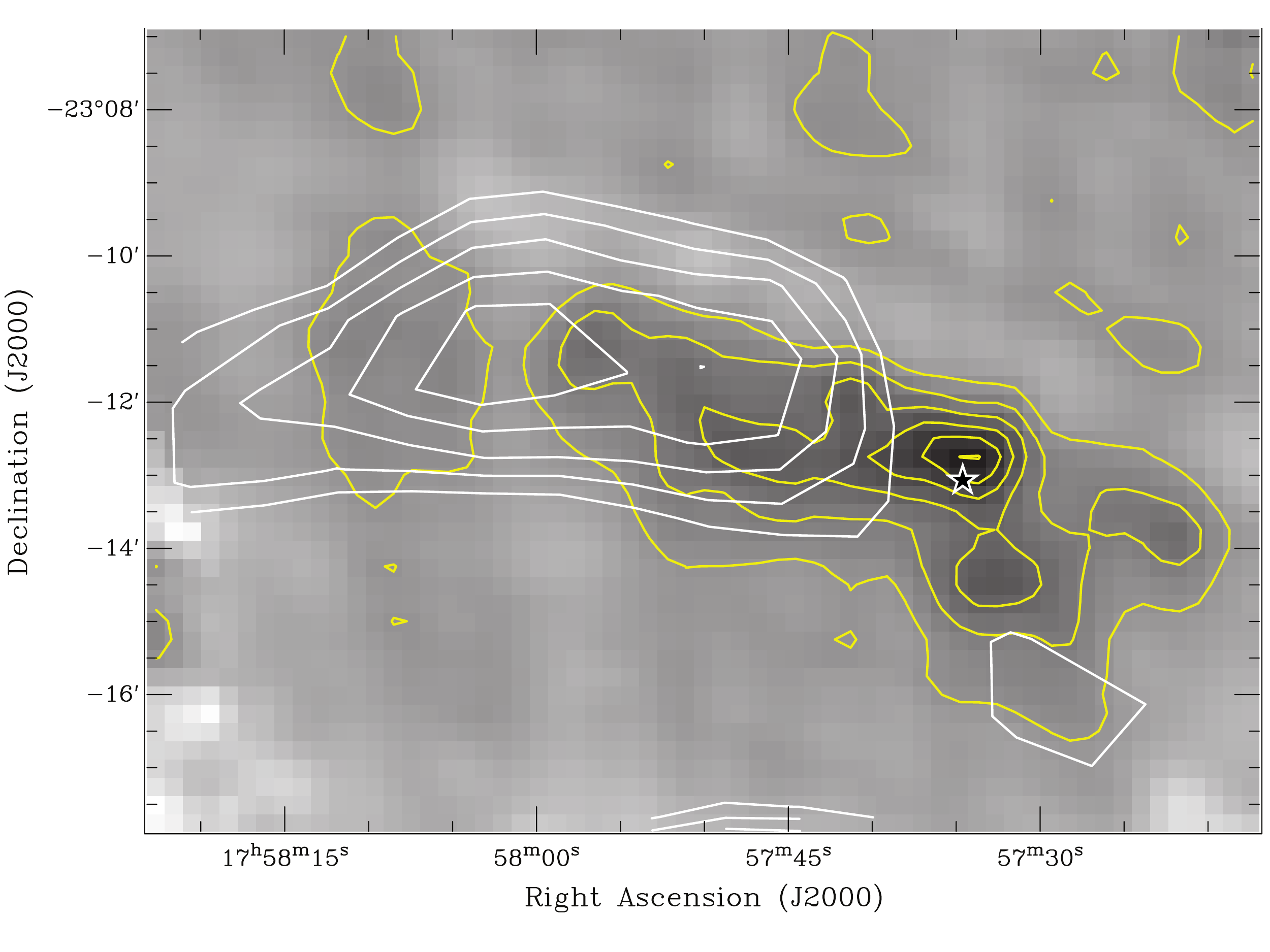}
\caption {Composite radio image of G6.31+0.54. Yellow contours are from the NVSS sky survey and white contours from the PMN. The yellow 
NVSS 20~cm contours  go from 0.001 to 0.01 Jy beam$^{-1}$ and the PMN 6~cm contours go from 0.01 to 0.03  Jy~beam$^{-1}$.  The underlying 
greyscale image is also NVSS.  The strongest NVSS yellow contours to the west are very close to the star  \mbox{TYC 6841-1665-1} that is marked with a white-black star symbol. However, the contours indicate the source is extended so it is possible there is a background source responsible rather than the star itself - most stars are radio quiet. 
Most of the G6.31+0.54 radio flux at this wavelength lies to the East of this star. }  
\label{PMN_NVSS}
\end{figure}

In addition to the 90, 20 and 11~cm observations noted by \citet{bro06}  using the NVSS data  \citet{Con98}, 
we made an additional search for G6.31+0.54 in other 
radio surveys. We found it was detected in the Parkes-MIT-NRAO \citep[PMN - see][]{cgw93} survey at 6~cm (4850~MHz) in the 
form of a partial shell (see Fig.~\ref{PMN_NVSS}) extended in the East-West direction at 11$\times$5.5~arcminutes 
(c.f. Fig.~2 of \citealt{bro06}). The strongest flux in the Western part was estimated to be
0.03 Jy~beam$^{-1}$. We could not estimate the radio flux of a partial shell at this wavelength due to the missing flux of extended objects in 
the PMN survey. A running median of 57~arcminutes was removed from the PMN scans  - see details in~\citet{cgw93} and also in \citet{stu05}.

\begin{figure}
\center
%Fig.2
\includegraphics[width=230pt, height=160pt, bb = 0 0 539 313 clip,]{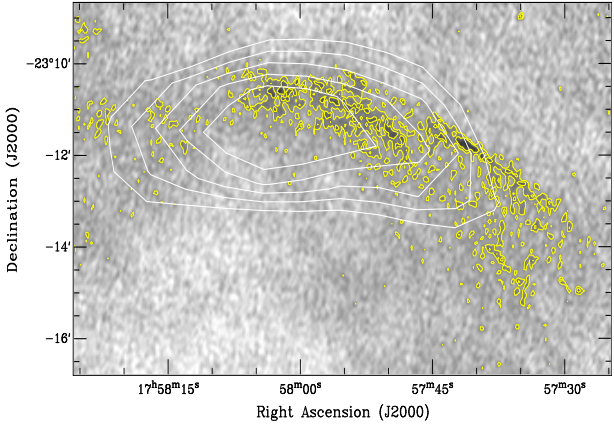}
\caption {G6.31+0.54 on Multi-Array Galactic Plane Imaging Survey (MAGPIS) 20~cm image. Again, white contours are from PMN 6~cm survey and  yellow 
from MAGPIS  going from     0.003 to 0.014  Jy~beam$^{-1}$.  The peak emission has match with peak emission on 20~cm NVSS image, but this MAGPIS 
image is with much better resolution ($\sim$ 5~arcsec). }  
\label{PMN_MAGPIS_G631}
\end{figure}

Fig.~\ref{PMN_NVSS} shows a composite radio image of G6.31+0.54. The yellow contours are from the NVSS sky survey while white 
contours give the PMN data.  The underlying greyscale image is the NVSS data. There is decent agreement but not a complete match between 
both wavelengths. This is probably because of the PMN survey's lower resolution ($\sim$5~arcminutes) 
compared to the NVSS of $\sim$45~arcseconds. The low NVSS surface brightness sensitivity can also influence visibility. The strongest 
NVSS yellow contours to the West are very close to the bright star  \mbox{TYC 6841-1665-1}. In the NVSS 
radio catalogue there is a source $\sim$23~arcseconds away that agrees with star position to within the NVSS positional uncertainties  \citep{Con98}. However,
the radio contours indicate the source is slightly extended and so despite the close positional co-incidence and stellar brightness the radio source could be a background object, recalling that most stars are radio quiet. Most of the G6.31+0.54 radio flux at this wavelength lies to the East of this star. \citet{bro06} used 20~cm data for a spectral 
index estimate, but from other surveys and not from the NVSS \citep[details in][]{bro06}, probably due to the bad quality imaging of this survey.\footnote{NVSS 
observations were performed with short ``snapshots"  giving bad quality imaging and sometimes highlighting  artefacts in complex regions.}  At 
20~cm G6.31+0.54 is also visible in the images the Multi-Array Galactic Plane Imaging Survey (MAGPIS) shown on 
Fig.~\ref{PMN_MAGPIS_G631} \citep[see details in][]{helf06}. It is clear that the definition of G6.31+0.54 in MAGPIS is much better compared to NVSS (Fig.\ref{PMN_NVSS}) due to the excellent resolution ($\sim$ 5~arcsec) of the MAGPIS survey.

\section{Infrared observations}

An extensive search for infrared counterparts to G6.31+0.54 was made across available infrared surveys. This included all wavelength bands from 
the MSX mission, \citet{price01},  the WISE  survey \citet{wri10}  and the GLIMPSE programs from the Spitzer observatory, \citet{ben03}. No trace of this remnant 
was found in any infrared wavelength with the same or similar morphological form as seen in the radio (and optical) data. In several MIR images  there does 
appear to be  Galactic dust emission in the region but without any obvious connection to G6.31+0.54. We also noticed  emission very close to the peak 
flux from the PMN 6~cm data that coincides with an infrared emitting star marked as IRAS~17550-2311. This is not to be confused with the peak in the 
NVSS data that is found in close proximity to a different, very bright star  \mbox{TYC6841-1665-1} that is itself unlikely to be the source of the radio emission. 

However, further investigations of the Spitzer mission MPSGAL data of the region  \citep{car09} does indicate a possible 
connection of 24$\mu$$m$ features with the radio emission of G6.31+0.54 at 20~cm where
%It is clear from  Fig.~\ref{MPSGAL} that 
an infrared filament exhibits a morphological structure similar to the optical/radio structure though this filament is only  $\sim$50~arseconds in length.  
This gives sufficient evidence to indicate that G6.31+0.5 may have infrared emission (at least at 24$\mu$m).

\section{Optical observations}
\subsection{Optical morphological structure}
Recent high resolution, high sensitivity \HA\ surveys give fresh impetus to optical detection of previously known radio SNRs and have 
uncovered many new remnants not previously noted in the extant radio surveys \citep[e.g.][]{stu08}. We present, for the first time, the 
optical counterpart to SNR candidate G6.31+0.53 from the SHS \HA\ survey of the Southern Galactic Plane  \citep{parker05}.  
The survey has arcsecond resolution and the \HA\ imagery permits detection of faint, extended \NII\ $+$ \HA\ emission down to a \HA\ surface 
brightness $\sim2\times10^{-17}$ erg ~cm$^{-2}$ s$^{-1}$ arcsec$^{-2}$ ($\sim3$~Rayleighs). This makes the survey sensitive to evolved, low
surface brightness nebulae \citep{Frew14} such as evolved SNRs (e.g \citealt{stu08}) and planetary nebulae (e.g. \citealt{parker06}).

\begin{figure}
\center
\includegraphics[width=230pt,height=170pt]{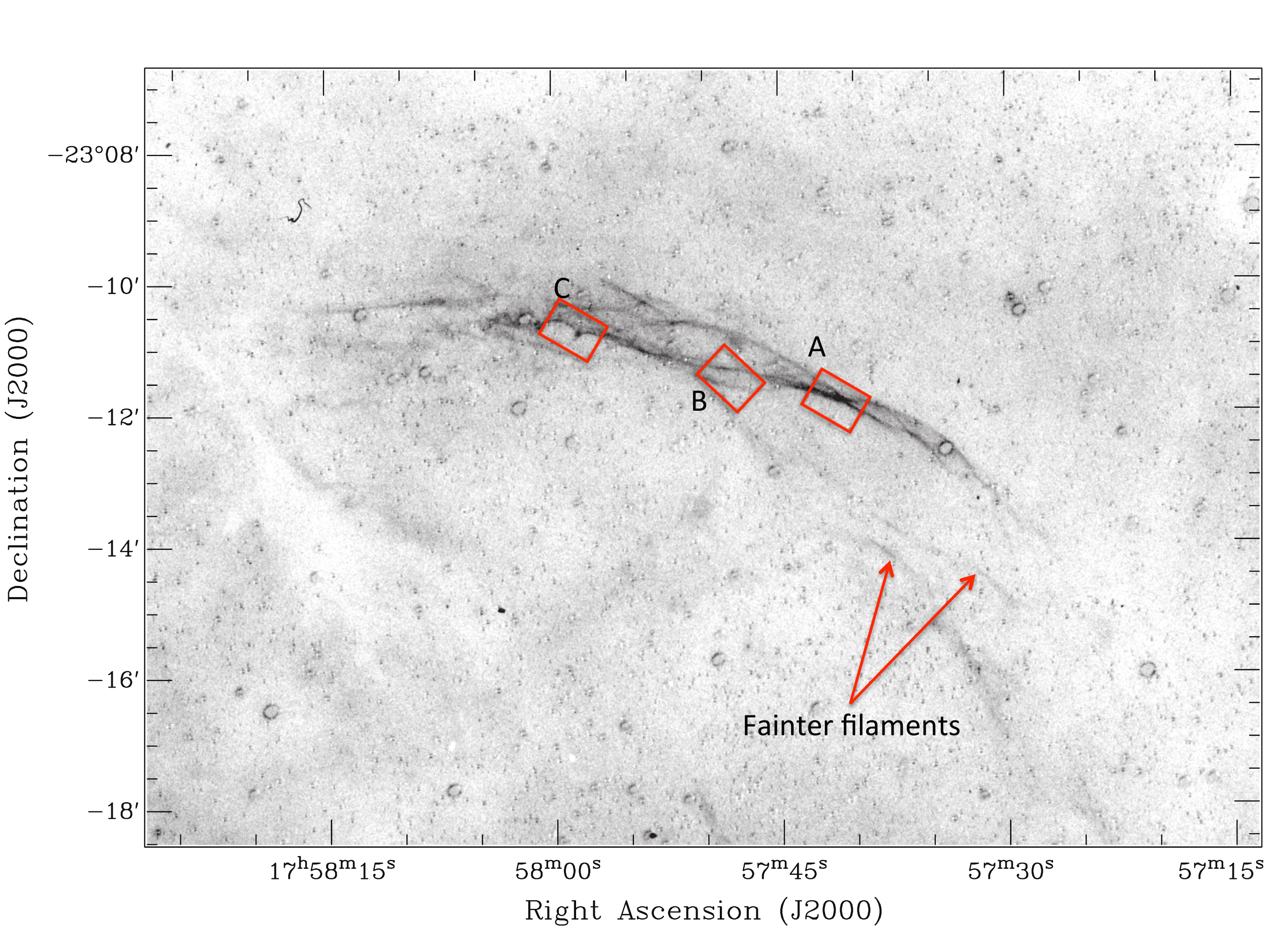}
\caption {A $12\times12$~arcminute quotient image of G6.31+0.54 made from the arcsecond resolution \HA\ and broad-band SR (red) images. Apart 
from the strong and clearly defined filamentary structure (partial shell)  of the front shock, some lower surface brightness filaments (arrowed) can be noticed 
south of main filaments of G6.31+0.54. These are difficult to see against the background in the \HA\ only image (see Discussion). The footprints of the 3 WiFeS positions A, B and C are also indicated.}
\label{quotient}
\end{figure}

To better reveal G6.31+0.54 we present a quotient image from dividing the \HA\  data by the matching broad-band short red (SR) equivalent.  
G6.31+0.54 has elements of a classical SNR optical form  (see Fig.~\ref{quotient}). It is visible as a partial shell some 10~arcminutes in 
extension with fine filaments that usually represent the remnant's shock front. A comparison with Fig.~\ref{PMN_NVSS} shows an excellent match 
with the NVSS emission which completely follows the strongest  \HA\  fine filaments,  despite  the low surface brightness of the NVSS  survey. A 
decent match with the low resolution image taken from the PMN survey is also seen across the main optical features. A series of fainter optical 
filaments (indicated) are seen extending from the middle of the main optical arc to the South-West which could also be related (refer to later discussion).

We also searched the digital version of the earlier epoch Palomar Observatory Sky Survey photographic atlas (POSS-I)  on the broad-band red 
exposures\footnote{See link~\nolinkurl{http://stdatu.stsci.edu/cgi-bin/dss_form}} for any trace of the fine filaments seen in the SHS  (Fig.~
\ref{quotient}) but could only find some very faint undefined 1-2~arcminute  nebulosity at the Western part of the main filamentary structures. This 
demonstrates the power of narrow-band \HA\ imaging to reveal SNR optical counterparts (\citealt{stu08}; 
\citealt{stu11}).  The  high sensitivity but low resolution (48~arcseconds/pixel) Southern \HA\ Sky Survey Atlas (SHASSA) of \citet{gaus01} was 
also examined where G6.31+0.54 is seen in a form similar to the PMN radio image at this low optical resolution. Fine 
filamentary structures cannot be resolved but as the SHASSA survey is carefully flux calibrated a peak \HA\ emission of 730~deciRayleighs (dR) was 
estimated from this data.

\subsection{Optical spectral observations}

The SNR \HA\ detection indicates decent spectra would be obtained from pointings along the bright filaments 
seen in Fig.~\ref{quotient}. Spectral observations were taken in June 2011 using the Wide-Field Spectrograph (WiFeS)  on the 2.3m 
telescope of Mount Stromlo and Siding Spring Observatory (MSSSO). This was to confirm the nature of this candidate via the observed emission 
lines and associated diagnostic line ratios (e.g. see Figs~7 \& 8 of Sabin et al., 2013). WiFeS is an image slicer and behaves as an integral field unit 
(IFU) providing spatial spectroscopy \citep[see details in][]{dop07}. WiFeS consists of $25\times1$~arcsecond wide adjacent slits which are each 
36~arcseconds in length to yield an effective $25\times36$~arcseconds on-sky footprint.  These footprints are also indicated in Fig.~\ref{quotient}. The chosen gratings gave a 
combined low resolution coverage across the separate blue and red arms of the spectrograph. After standard processing with the WiFeS data reduction pipeline we chose suitable 
apertures from the WifeS footprint to extract summed 1-D spectra. These extracted apertures provide an integrated \mbox{1-D} spectrum of all the flux within 
the selected region. Our choices were based both on selecting regions where the flux is  strongest but also making sure to sample along the structure of the partial SNR shell. 

Table~1 gives June 2011 observation log of SNR G6.31+0.54. The listed R.A. and Dec. are for the positions of the centre of each WiFeS 
footprint A, B and C. The blue and red arms of the spectrograph gave a combined wavelength coverage from  3200 and 7000\AA\ using a medium resolution 
grating of 708~lines mm$^{-1}$ for the blue arm and a higher resolution 1200~lines mm$^{-1}$ gratings for the red arm for improved kinematic 
capability and to resolve the key [SII] doublet used to estimate electron densities. 

Here we present 1-D flux calibrated red spectra extracted from selected apertures for two of the three IFU positions A and B from 
Table~1. For both nights we used observations of the EG274 photometric standard star to assist with determination of flux calibrated spectra. 

\begin{table}
  \setcounter{table}{0}
  \centering \setlength{\tabcolsep}{1.6mm}
  \scriptsize{\caption{Observing log for the MSSSO 2.3m telescope WiFeS IFU exposures of SNR G6.31+0.54  taken in June 2011.
  Grating B3000 (708 lines mm$^{-1}$) was used for the blue  arm and R7000 (1200 lines mm$^{-1}$) for the red arm of the spectrograph.}
 \setlength{\tabcolsep}{1pt}
 \begin{tabular}{cccccc}
 \hline\noalign{\smallskip}
 WiFes &Date  & Exposure &Spectral& &  \\
  position && (sec)&~range (\AA)&RA&Dec.\\
 &&& blue + red&&\\
 \hline\noalign{\smallskip}
 &&&&$~~~~^h$~~$^m$~$~^s$&~~~~~\D~~~'~~~''\\[-3pt]
 A&02/06/2011&1500&3200--7000&17~57~41&--23~11~45\\
 B&02/06/2011&1500&3200--7000&17~57~49&--23~11~20\\
 C&03/06/2011&1200&3200--7000&17~57~57&--23~10~50\\
 \hline
 \end{tabular}}
 \vspace{-5pt}
 \begin{flushleft}
 \tiny $^a$The rms dispersion error (in \AA) was 0.08 and  the relative percentage error in the flux estimate  was
 $\sim$12\%.\\
 \end{flushleft}
 \label{table1}
 \end{table}
 
For position ``A" from Table.1 we present red spectra in Fig.~\ref{A_red} where 
the left side shows a full slice from the WiFeS datacube centered on the strongest line in the optical spectrum of \NII\ at 6583\AA. The selected 
aperture from footprint ``A" was  large, 22$\times $30~arcseconds which represents 73\% of the entire WiFeS footprint at this position. This aperture 
was summed through all the cube to give the 1-D spectrum shown on the right side of Fig.~\ref{A_red}. 

The usual \HA, \NII\ 6548,~6583 and \SII\ 6716, 6731\AA\ emission lines are widely used to identify the nature of the observed nebulae based on 
criteria established by \citet{Fesen85} and refined by \citet{Frew10}. All these major, red emission lines, typical for older supernova remnants, are 
present, i.e. \NII\ at 6548 and 6583\AA, \HA\ and \SII\ at 6717 and 6731\AA. 
The \NII\ lines are stronger than \HA\ which immediately rules out possible confusion with a \HII\ region (e.g. \citealt{Frew10}).  While strong \NII\ 
emission is also seen in both planetary nebulae and massive star ejecta \citep{Frew14b}, the resolved \SII\ lines in summation are stronger than \HA\  
showing we are well in to the shock regime typically seen in SNR optical spectra. A  \SII\ 6716 + \SII\ 6731/\HA\  ratio of $>$ 0.4$\sim$0.5 (see  
\citealt{Fesen85}) is a main discriminator for the presence of shocked conditions of the energetic and turbulent regions of SNRs. Such values are very 
rarely found in \HII\ regions or planetary nebulae. The value found here is $\sim$1.2 and is obtained from using the integrated fluxes obtained from 
Gaussian fits to the line profiles. Furthermore, the observed \SII\  6717/6731\AA\ ratio is used  to determine electron densities. For most SNRs this is 
likely to be in the low-density limit of $\sim$1.4 and indeed the measured value 
from position A is $\sim1.3\pm0.1$ (see Table~2).

\begin{figure}
\center
\includegraphics[width=245pt,height=178pt, bb = 0 0 403 270, clip]{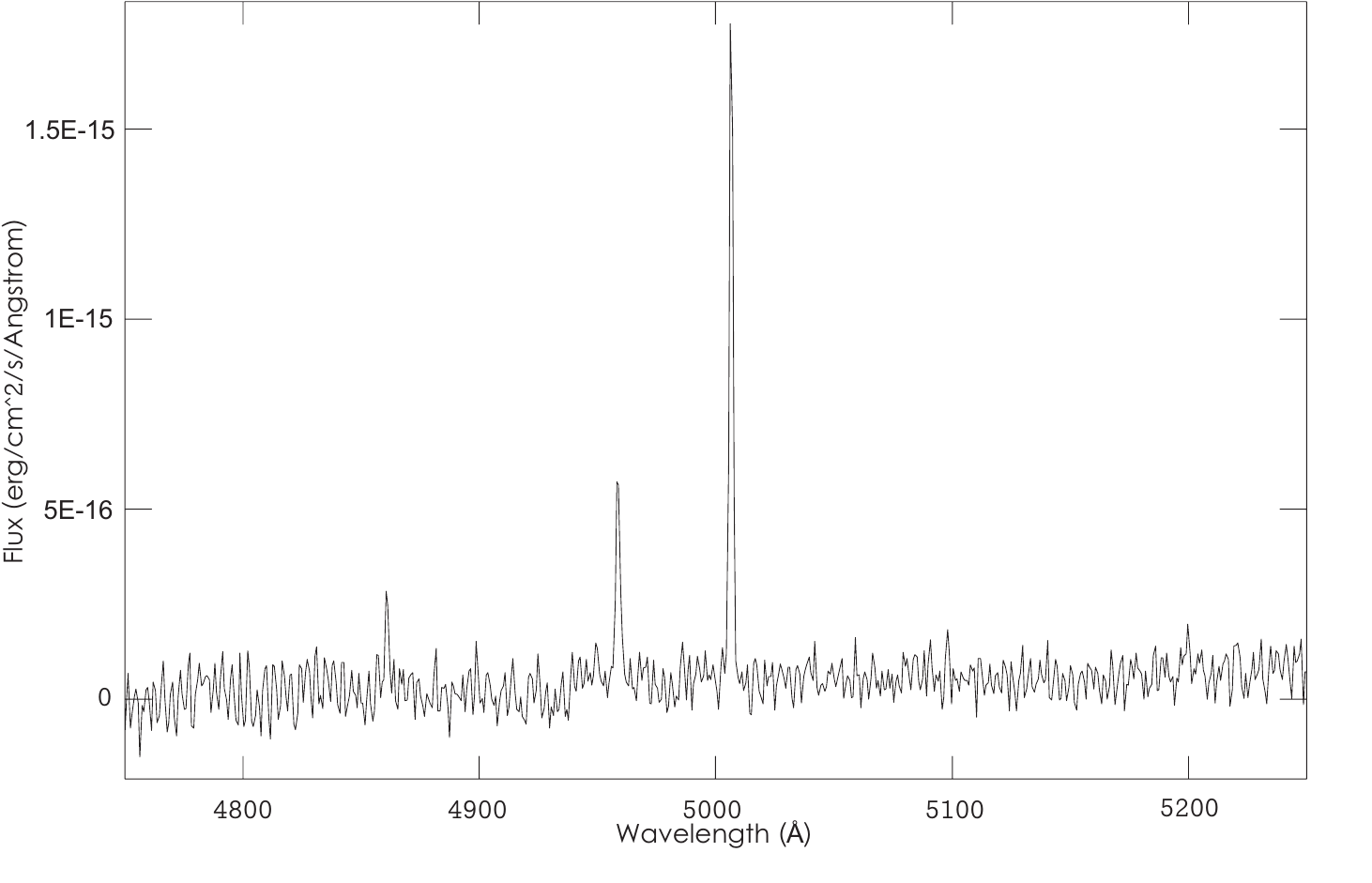}
\caption {1-D flux calibrated  position ``A"  blue spectrum of WiFeS footprint for G6.31+0.54 taken through a 5$\times$23~arcsecond aperture.  The 
\HB\ line and \OIII\ lines  at 4959 and 5007\AA\  are detected (see also Table~2) with the \OIII\ lines giving a ratio of 2.6 slightly lower than the 
standard ratio due to extinction. }
\label{A_blue}
\end{figure}

\begin{figure}
\center
\includegraphics[width=245pt,height=190pt]{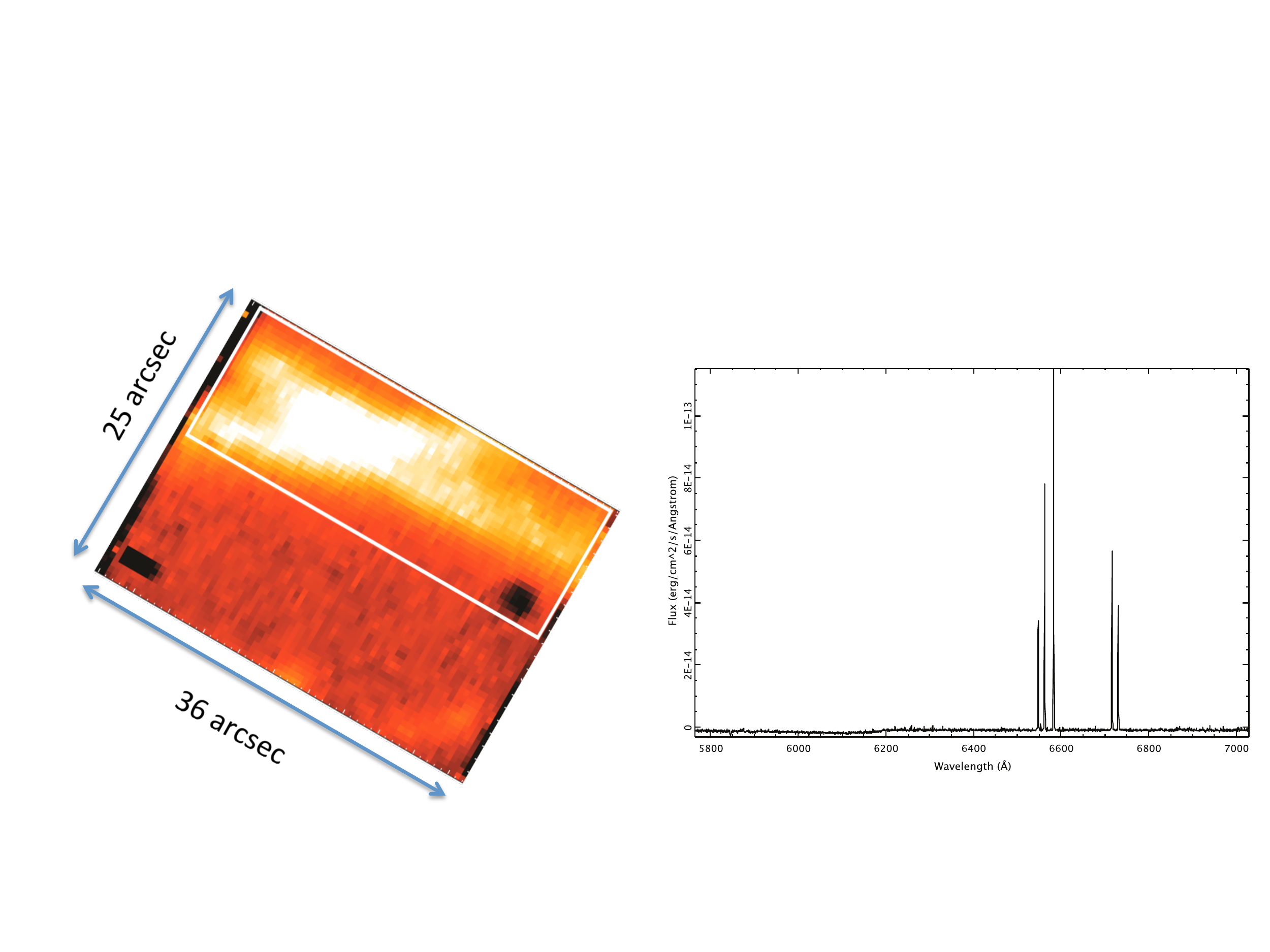}
\caption {The position ``A" WiFes footprint for G6.31+0.54 (position given in Table~1)  as oriented on the sky with respect to the H$\alpha$ quotient image in Fig.~\ref{quotient} with the image slice centered at \NII\ 6583\AA\ from the WiFeS datacube. The white rectangle gives the size of the extracted aperture (22$\times$30~arcseconds) used for providing the 
flux calibrated 1-D spectrum shown on the right side of the image. This red spectrum shows extremely strong \NII\ at 6583\AA\  relative to \HA. The \SII\ lines at  6717 and 6731\AA\ are also very strong giving a ratio of  \SII\ against \HA\ of $\sim$1.2 that confirms shock excitation typical for supernova remnants.}
\label{A_red}
\end{figure}

Due to extinction all blue footprints had low S/N but a reasonable  blue spectrum was possible from the brightest ``A" filament (see Fig~\ref{A_blue}) 
that shows classical lines of  \HB\ and. \OIII\ at 4959 and 5007\AA.

\begin{figure}
 \center
\includegraphics[width=240pt,height=190pt]{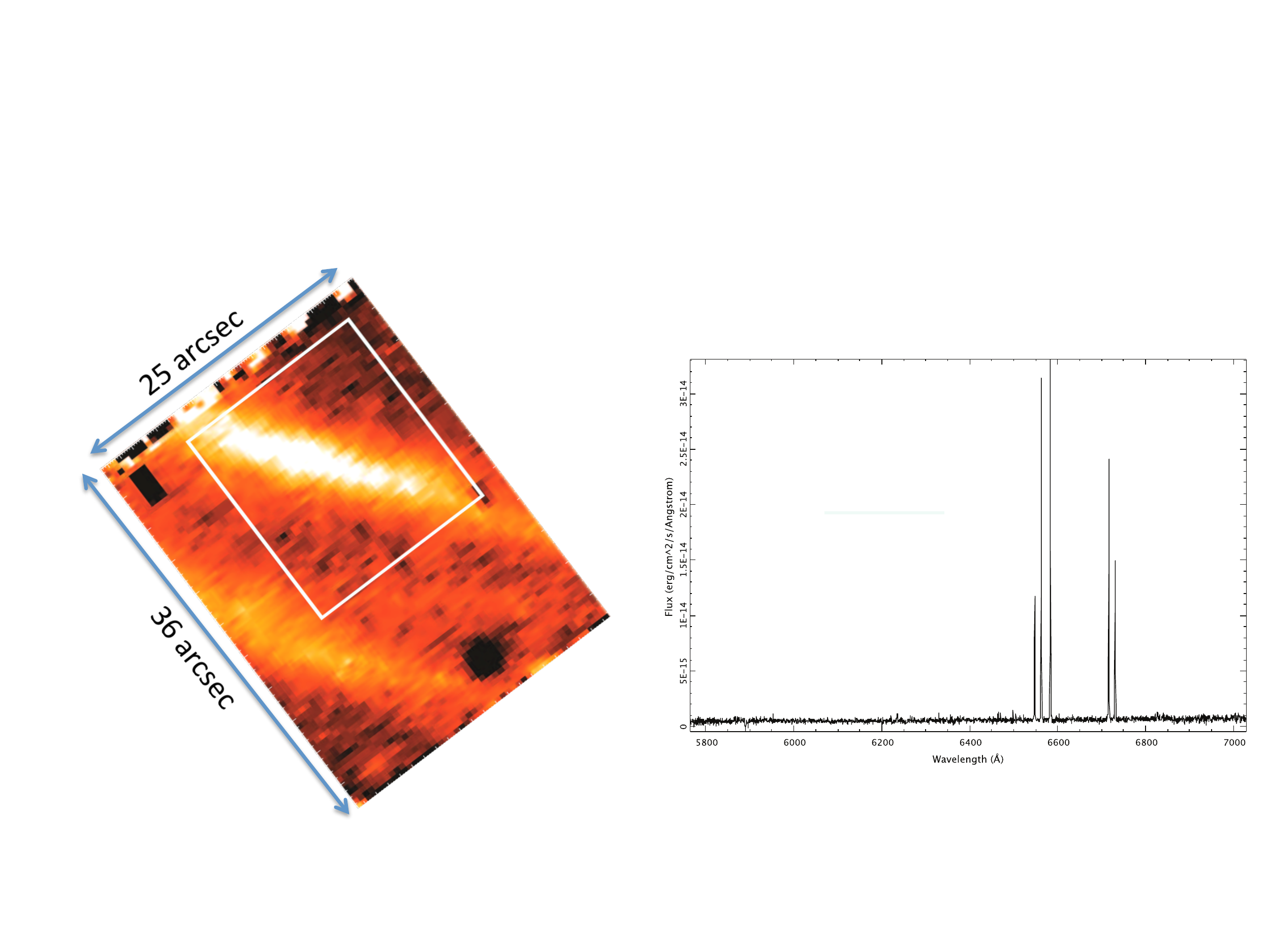}
\caption {``B" position WiFes footprint for G6.31+0.54 (refer Table~1)  as oriented on the sky 
with respect to the H$\alpha$ quotient image in Fig.~\ref{quotient}  with the image slice centered  
at \NII\ 6583\AA\ from the datacube.  The white rectangle shows the 16$\times$20~arcsecond aperture used to extract the 1-D spectrum 
shown at the image right side. The spectrum has strong \NII\ lines at 6548 and 6583\AA relative to \HA\ and very strong \SII\  at  6717 and 6731\AA relative to \HA\ with a ratio of $\sim1.2$ obtained from using the integrated fluxes obtained from Gaussian fits to the line profiles.}
 \label{B_red}
 \end{figure}

Similarly, the integrated 1-D red spectrum at position B is presented in Fig.~\ref{B_red} where again the WiFeS 2-D cube slice seen at  \NII\ 6583\AA\ 
is shown. The selected aperture was 16$\times$20~arcseconds in size chosen to sample the most intense section of the filament. The 1-D spectrum 
on the right of the figure shows the same lines as seen extracted from 
position A  but with slightly different observed ratios.  The equivalent blue data gave weak detections of  \HB\ and \OIII\ at 5007\AA~due to extinction 
which is why only the ratio of \OIII\ against \HB\ is given in Table~2.

Output from WiFeS position C shows the same lines seen 
in 1-D spectral extractions from  WiFeS footprints A and B with similar line ratios though the spectrum extraction aperture was only a 3~arcsecond region from the brightest emission seen in the WiFeS cube and located some 11~arcseconds North from the R.A. and Dec. position given in the Table~1 was used.  
An exposure  time of 1200~sec was used for position C and 1500~sec for positions A and B (refer Table~1). 
The  blue spectrum is of too low S/N from this modest aperture and did not 
show any clearly detected spectral lines though the observing log shows clear nights and good seeing.  Despite observations of spectrophotometric 
standard stars a reliable flux calibration for the combined blue and red spectra was not achieved due to the different blue and red gratings and 
different CCDs used for the independent blue and red spectrographs arms. Consequently it was not possible to estimate a reliable Balmer decrement 
and hence extinction in the direction of G6.31+0.54.

\begin{figure}
 \center
 \includegraphics[width=230pt,height=220pt, bb = 215 13 791  552, clip]{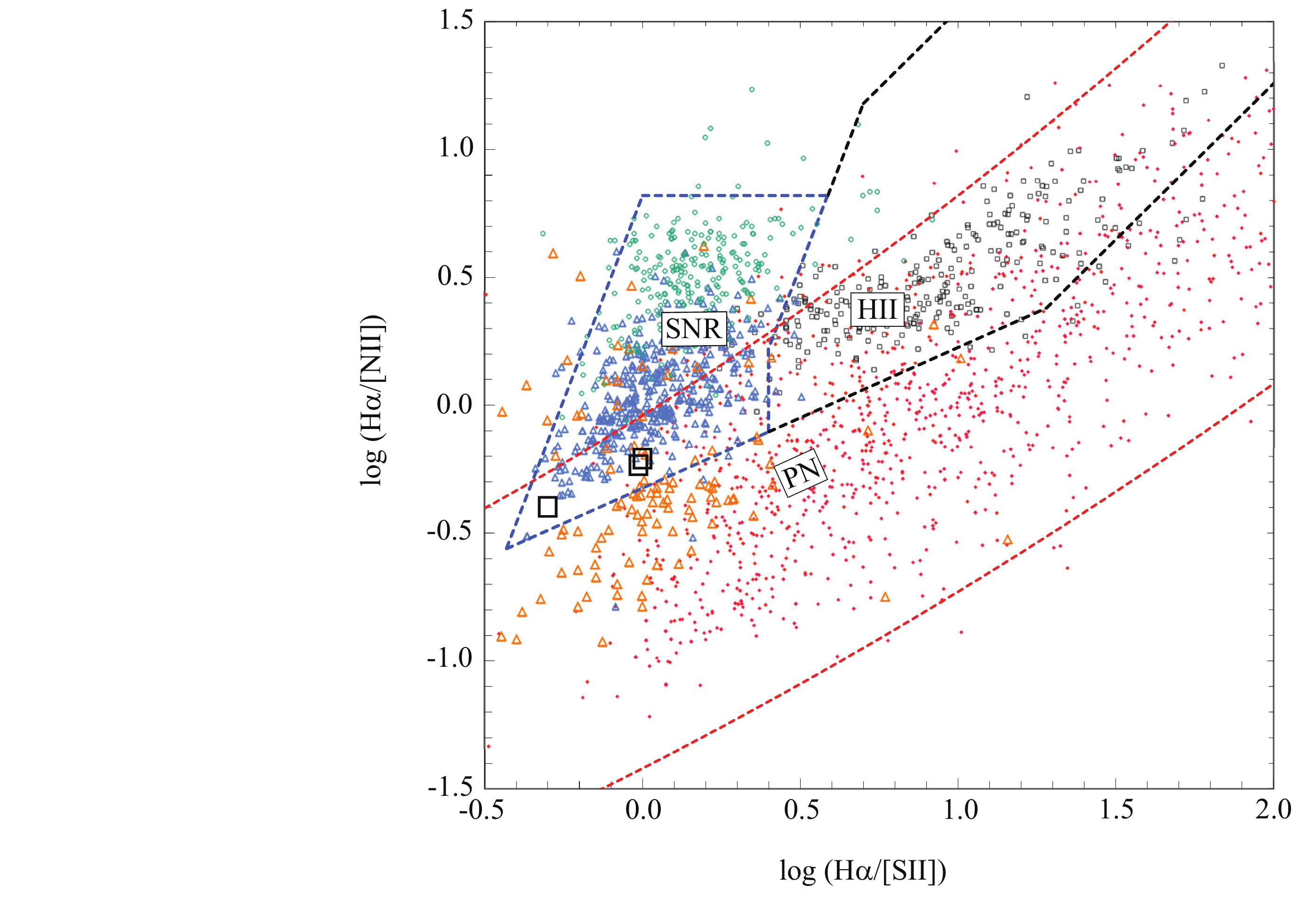}
 \caption {The ``SMB"  \HA/\NII\ versus \HA/\SII\  diagnostic diagram from \citet{sab13}. Galactic PNe are plotted as red dots, HII regions as black open squares, evolved Galactic SNRs as blue open triangles, young Galactic SNRs as orange triangles and the low-metallicity Magellanic Cloud SNRs as green triangles. The PN region is bounded by the red dashed curves, the evolved SNR field (of relevance here) by the dashed blue lines and the HII region domain by the dashed black lines. The black squares inside the SNR area show the data 
 positions of the A, B and C spectrographic observations.  It is clear that the data from G6.31+0.54 for all three positions fall well within the SNR 
 defined loci in the plot, clearly demonstrating their likely SNR nature.}
 \label{smb-bpt}
\end{figure}

Values for selected line ratios obtained from positions \mbox{A, B and C} can be plotted on the \mbox{so-called} ``SMB" \HA/\NII\ versus \HA/\SII\ 
diagnostic diagram of \citet{sabb77} or the \citet{Bald81} BPT \OIII\ 5007/\HB\ versus \SII/\HA\ equivalent. Both can be used to help discriminate 
between various classes of emission nebulae including SNRs, \HII\  regions and PNe. These useful diagrams have been updated by our group (e.g. 
see \citealt{Frew10} and \citealt{sab13}) to include vetted spectroscopic data from the literature, supplemented with our own flux-calibrated 
spectroscopy for a large sample of objects of various kinds as reported in the HASH research platform (e.g \citealt{parker16}). These new data-points 
and their integrity have  provided a better definition of the overall loci for the different object types  with specific zones indicated.  
These SMB and BPT diagrams are presented in Fig.~\ref{smb-bpt} with the larger black squares  gives the data obtained from G6.31+0.54 positions A, B and C respectively. They all fall well within the SNR defined loci in this plot showing their highly likely SNR nature.  A few PNe do seem to fall in the SNR delineated region of the SMB diagram but they are a small minority. Any even conservative physical extent of this nebulae is $\sim10\times$ bigger than for even the largest PNe ever observed so there is no confusion.

 \begin{table*}
    \centering 
  \scriptsize{
  \caption{The measured ratios of the observed emission line intensities$^\dagger$ for G6.31+0.54 taking \HA=100 and \HB=100 for all three WiFeS 
spectrograph IFU positions from  Table.1. The blue and red line ratios are taken independently, e.g. compared separately according to \HB\ 
and \HA\ Balmer lines due to the different spectrograph arms used for  blue and red part of the spectrum and different apertures sizes used for the 
extraction of blue and red spectrum. }

  \setlength{\tabcolsep}{1pt}
  \begin{tabular}{@{\extracolsep{4pt}}ccccccccccccccc}    %bilo {4pt}umjesto \filll
  \hline\noalign{\smallskip}
  Date&Wifes  & \HB &\OIII &\OIII &\NII &\HA&\NII&\SII&\SII&\NII/\HA&\SII/\HA&\SII & radial velocity&electron density\\
  & position &&4959\AA &5007\AA & 6548\AA&&6583\AA &6717\AA &6731\AA &&&6717/6731\AA &(\ks)&(cm$^{3}$) \\
  \hline\noalign{\smallskip}
 02/06/2011&A&100$^a$&267&690&42&100$^a$&140&70&52&1.8&1.2&1.3 &-12.2$\pm$1.4&$\sim$10$^{2}$\\
  02/06/2011&B&100$^b$&--&125&35&100$^b$&113&70&48&1.4&1.2&1.5&-3.6$\pm$2.8&LDL \\
  03/06/2011&C&--&--&--&66&100$^c$&196&112&85&2.6&2.0&1.3&-22.2$\pm$3.3&$\sim$10$^{2}$\\
  \hline
  \end{tabular}
  }
  \vspace{-5pt}
  \begin{flushleft}
\hspace{0.7cm} \tiny $^a$\HA\ flux = 1.16$\times$10$^{-13}$ \& \HB\ flux = 5.16$\times$10$^{-16}$; $^b$\HA\
flux = 4.43$\times$10$^{-14}$ \& \HB\ flux = 1.91$\times$10$^{-16}$; $^c$\HA\ flux = 2.03$\times$10$^{-15}$ in units of erg cm$^2$ s$^{-1}$ \AA\
$^{-1}$\\  \hspace{0.7cm}\tiny $^\dagger$Note the rms wavelength dispersion error from the arc calibrations was 0.08\AA while the relative 
percentage error in the flux determination from \\
  \hspace{0.7cm}\tiny the calibration using the brightest lines was estimated as $\sim$12\%.\\
\end{flushleft}
  \label{table2}
  \end{table*}

\section{X-ray Observations}

\begin{figure}
\center
\includegraphics[width=230pt,height=220pt, bb = 0 44 538 500, clip]{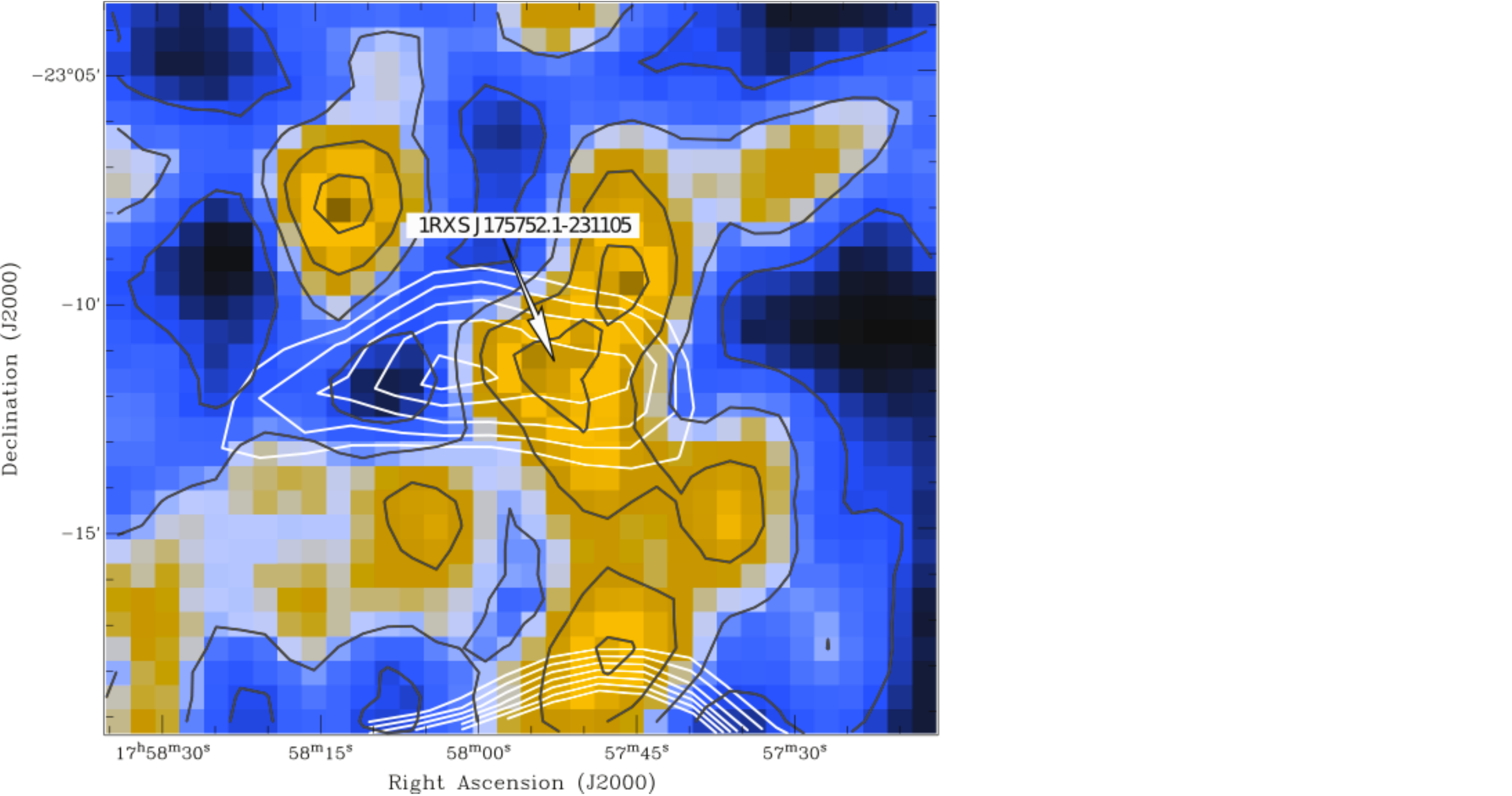}
\caption {The ROSAT X-ray data for G6.31+0.54 from the PSPC image  from
0.1 to 2.4~keV. The image has been binned so one pixel is 32~arcseconds and smoothed using a Gaussian filter with $\sigma$=2. The 
ROSAT source \mbox{1RXS J175752.1-231105} is marked as the most probable association with G6.31+0.54. The white contours are 
PMN survey data from 0.01 to 0.04 Jy~beam$^{-1}$) and shows a good positional overlap between the radio, X-ray and optical emission for this object.}
\label{xray}
\end{figure}

A search for X-ray sources around G6.31+0.54 revealed  ROSAT object \mbox{1RXS J175752.1-231105} in the 
All-Sky Faint  Source Catalogue (RASS-FSC) \citep{vog2000}~\footnote{see also link http://www.xray.mpe.mpg.de/rosat/survey/rass-fsc/} that has a 
match with the regions optical and radio emission in Fig.~\ref{xray}. This figure, with the ROSAT source  \mbox{1RXS 
J175752.1-231105} indicated, is for the range  0.1 to 2.4~keV extracted and smoothed with a $\sigma$ of 2-bins 
on the basis of the region's ROSAT All Sky Survey image (using \underline{ }bas.fits file). This ``basic" ROSAT file contains all essential calibrated 
science data such as photon event list with time of arrival, length of observations, sky and detector positions, corrected amplitude for each photon etc.  
The RASS-FSC does not give the diameter of this source but we estimate it is around 2~arcminutes.  The source is in the group of faint objects with a 
likelihood detection of 12  where a likelihood scale of at least 7 indicates detection of at least 6 source photons. The ROSAT catalogue entry for this 
source gives only a coarse description of the spectrum for this object, or hardness ratio (HR1), defined as the ratio of lower energy 
end of the spectrum (soft end e.g. 0.1 to 0.4~keV)  and the high-energy spectrum (hard end e.g. 0.5 to 2.0~keV). The listed hardness value,
actually a ratio of counts of two different energy bands, is 1.0~$\pm{0.2}$ for \mbox{1RXS J175752.1-231105} telling us we have very hard emission 
of hot X-ray gas in this region which is common to young SNRs and in cases where the ISM is heated by a strong shock.  Although the observed 
hardness ratio is a rough estimate, it gives basic spectral  properties of the object. Despite good positional agreement no firm association of the X-ray 
source with the SNR can be made at this time.
%% Joe can we say a little more about the hard X-ray flux expected from old remannt?

\section{Discussion and Conclusions}

If we compare the optical image of G6.31+0.54 (Fig.~\ref{quotient}) with the NVSS~20~cm radio image (yellow contours in Fig.~\ref{PMN_NVSS})  
we see an excellent positional match. The radio contours at 6~cm from the PMN survey are shifted a little towards the East compared with the NVSS 
and the optical, but this is likely due to the low angular resolution of the PMN  survey  ($\sim$5~arcminutes). The peak emission at 6~cm is clearly 
defined while in the NVSS the peak emission is close to the position of the bright star \mbox{TYC 6841-1665-1} 
but the slightly extended nature of the radio emission here indicates a likely background source rather than any direct association with the star itself. 

The full extent of the multi-wavelength emission distribution of G6.31+0.54  is unclear but what is seen in all radio and optical images is a  partial shell  
$\sim$10~arcminutes in extent that morphologically represents the shock front of an evolved Galactic SNR.  If we assume the observed feature is 
part of an arc of a circular shell then the conservative approximate diameter would be about 20~arcminutes obtained by subjective fitting of a circle to the 
observed arc structure. The clear detection of radio emission at several frequencies and the resultant observed negative radio spectral index 
shows that it is still a radio active object and likely from an SNR but definitely not in the group of radio quiet remnants \citep[see ][]{stu07}.  
That it is an older evolved remnant is corroborated by our optical spectra (see Fig.~\ref{A_red},\ref{B_red}) where the observed ratios of 
the forbidden lines of  \NII, \HA\ and \SII\ do not vary too much across the visible remnant (see also Table~2). This is not the  case with young 
remnants. Also, the ratio of \SII/\HA\ $>>$ 0.4  (in our case it is 1.2,~1.2~and~2.0 for the three sampled positions) is an established  metric  for objects 
with shocks, e.g. SNRs or, in some cases Wolf-Rayet star nebula \citep[see][]{stu11}. Together, all these data  tell us that G6.31+0.54 can be 
classified as a coming from a newly confirmed Galactic SNR whose full extent is unclear.

Very faint filamentary emission features appear to start in the middle of the main emission arc of G6.31+0.54 and extend South-West for 
several arcminutes (see Fig.~\ref{quotient}).  The nature of these faint filaments (arrowed in the figure) 
and their possible association with this SNR is not certain but 
could indicate continuation of the optical shell. Connection with the known Galactic SNR G6.4-0.1 (or W28) 
10~arcminutes South-East from G6.31+0.54  (see middle image of top panel in Fig.~2 in Brogan et al., 2006) is possible but seems unlikely given 
the clear separation.  Only future detailed study can offer a firm answer. G6.31+0.54 is also quite close to another SNR G6.10+0.53 but given the 
13~arcminute separation in declination they appear unrelated (see see Fig.~3 in \citet{stu11}).

From the optical images presented in this work it is apparent that the filaments of  G6.31+0.54  are on the rim (e.g. part of the rim) of the partial SNR shell 
and seen edge-on (tangentially).  In this case \citep[see also][]{bou04}  we assume that radial velocities  characterise the majority motion of the 
expanding system. Hence, from the estimated radial velocities we can calculate a crude kinematic distance to G6.31+0.54. For radial velocities for all 
three slit positions from Table~2 we measured shifts of the brightest well resolved lines of \NII, \HA\ and 
\SII\ and estimated an average radial velocity of \mbox{-12.7 \k}. Taking the Sun's radial distance from the Galactic center of \mbox{8.5~kpc}  gives an 
estimated distance of \mbox{$\sim $4.5~kpc} and therefore a tentative diameter of $\sim$26pc. Further spectra across the remnant would provide 
more accurate averaged velocities. We have estimated radial velocities from \mbox{-3.6\k$\pm$2.8\k} for position B, to 
\mbox{-22.2\k$\pm$3.3\k} for position C  with position A giving an intermediate value of \mbox{-12.2\k$\pm$3.1\k}.  
Due to these observed variations in radial velocities the distance of \mbox{$\sim$4.5~kpc} could have an error of up to \mbox{50\%}.

In looking for evidence for SNR evolution over time and as a possible independent test our distance estimate we compared available optical images  
(in red light)  from the early epoch Palomar Observatory  Sky Survey (POSS~I) and the much later SHS data to see if 
there was any discernible shift (expansion) in filament positions. Despite the 52 year time difference we do not see any evidence for change, perhaps not 
surprising given the small shifts of $\sim$0.2-0.3~arsceconds that might be expected with the supposed distance of this remnant and the coarser, earlier 
emulsions.

The spectra in Fig.~\ref{B_red} shows \NII\  6583\AA\  stronger than \HA\  but in  Figs.~\ref{B_red} and for the spectrum of pointing C this line is 
much stronger. This is not common for evolved SNRs that have chemical compositions dominated by swept-up 
ISM and show an abundance gradient across the Galaxy (e.g. \citealt{bin82}; \citealt{Fesen85}). Such  enrichment can arise due to 
localised nitrogen variability and interaction with the ISM (see  \citealt{stu09}~or~\citealt{stu07}). This variation in nitrogen 
abundance is more often seen in younger remnants. One more possibility is that strong \NII\ emission is due to a somewhat supersolar 
metallicity of the ISM closer to Galactic centre. 

In the blue spectra, none shows \OII\ at 3727\AA\ (common in evolved SNRs) which, according 
to \citet{ray79}  represents slower moving shocks. Lack of detection is more likely due to high extinction affecting the data in the 
far blue. From the blue spectra at positions A and B (see Table~2) the \OIII\ lines at 5007 and 4956\AA\  are clearly seen telling us that the 
shock velocity of G6.31+0.54 is likely to be over \mbox{80~\k}  ~\citep{ray79}. We used \citet{all08} to calculate the shock speed for the A and B IFU 
positions from Table~2. Strong lines of \SII\ at 6717 and 6731\AA\  
relative to \HB\ were used and very close values for both lines estimated, giving averaged shock velocity estimates of \mbox{$\sim$240~\k} for these 
two positions under the model of solar abundance \mbox{$n=1.0$~\cm3}  and pre-shock  magnetic field \mbox{$B=3.23$~$\mu G$}  \citep[see details 
in][]{all08}. If the observed filament is off the perpendicular to the line of sight by an angle $\theta$, then the observed radial velocity will be dominated 
by the filament peculiar velocity (240~\k $\times$ Sin($\theta$) is much greater than the systemic velocity of the remnant). In this case the kinematic 
distance estimate would be unreliable. For the same calculations we also tried to use the \NII\ ~line at 6583\AA\ relative to \HB\ but the subsequent 
estimated velocity was lower by some \mbox{50~\k}, most  probably due to the  enhanced \NII\ line in which case it is not a real physical characteristic of 
the source.

We have presented the first optical detection of SNR G6.31+0.54 in  \HA\  which exhibits typical filamentary emission structures common to SNRs  and 
that overlaps substantially with the previous radio data. We also presented the first optical spectra from several locations across the optical 
partial shell that confirm shock excitation typical of evolved SNRs. Diagnostic emission line diagrams such as the SMB  \HA/\NII\ versus \HA/\SII\  plot 
(e.g. Frew \& Parker 2010, Sabin et  al., 2013) places these spectra squarely in the 
locii of SNRs. We also identified this object in other radio surveys at different frequencies such as the PMN survey at 6~cm and NVSS at 
20~cm.   Finally, there is a ROSAT X-ray  
source \mbox{1RXS J175752.1-231105} that may be connected with G6.31+0.54 given the strong positional co-incidence.  Taking  the 
overall body of current observational evidence  we have compiled with the earlier determination of a negative radio spectral index representative of 
shell SNRs, G6.31+0.54  should be classified as coming from a bona-fide Galactic SNR. Further, deeper observations with different radio frequencies to better define the radio spectral index and form would be useful together with  more high dispersion optical spectroscopy and better X-ray data to 
confirm the possible association of \mbox{1RXS J175752.1-231105} with this SNR.   

\section{Acknowledgements}
We gratefully thank an anonymous referee for diligent and careful work that has considerably improved the paper.
We thank the Mount Stromlo and Siding Spring  Observatory Time Allocation Committee for time for the spectroscopic follow-up. 
MS is thankful to the Physics Department and Laboratory for Space Research of The University of Hong Kong, for a senior 
research assistantship to work on this project. We thank Prof. Miroslav Filipovi\'c for helpful comments during work on this paper.


\begin{thebibliography}{}

\bibitem[\protect\citeauthoryear{Allen et al.,}{2008}]{all08}
Allen M.G., Groves B. A., Dopita M. A., Sutherland R. S., Kewley L. J., 2008, ApJS, 178, 20

\bibitem[\protect\citeauthoryear{Baldwin, Phillips \& Terlevich}{1981}]{Bald81}
Baldwin J. A., Phillips M. M., Terlevich R., 1981, PASP, 93, 5

\bibitem[\protect\citeauthoryear{Benjamin et al.,}{2003}]{ben03}
Benjamin  R. A. et al., 2003, PASP, 115, 953

\bibitem[\protect\citeauthoryear{Binette et al.,}{1982}]{bin82}
Binette L., Dopita M. A.,  Dodorico S., Benvenuti P., 1982, A\&A,  115, 315

\bibitem[\protect\citeauthoryear{Boumis et al.,}{2004}]{bou04}
Boumis P., Meaburn  J.,  L\'opez  J.A., Mavromatakis  F., Redman  P.M., Harman D.J., Goudis  C.D., 2004, A\&A , 424, 583

\bibitem[\protect\citeauthoryear{Boumis et al.,}{2005}]{bou05}
Boumis P., Mavromatakis F., Xilouris E. M., Alikakos J., Redman M. P., Goudis C. D., 2005, A\&A, 443, 175

\bibitem[\protect\citeauthoryear{Boumis et al.,}{2008}]{bou08}
Boumis P., Alikakos J., Christopoulou P. E., Mavromatakis F., Xilouris E. M., Goudis C. D., 2008, A\&A, 481, 705

\bibitem[\protect\citeauthoryear{Boumis et al.,}{2009}]{bou09}
Boumis P., Xilouris E. M., Alikakos J., Christopoulou P. E., Mavromatakis F., Katsiyannis A. C., Goudis C. D., 2009, A\&A, 499, 789

\bibitem[\protect\citeauthoryear{Brogan et al.,}{2006}]{bro06}
Brogan C.L.,  Gelfand J.D.,  Gaensler B.M., Kassim N.E.,  Lazio T.J.W., 2006, ApJ, 639, L25

\bibitem[\protect\citeauthoryear{Cappellaro,  Evans \& Turatto }{1999}]{capp99}
Cappellaro E., Evans R.,  Turatto, M. 1999, A\&A, 351, 459

\bibitem[\protect\citeauthoryear{Carey at al., }{2009}]{car09}
Carey S. J. et al.,  2009, PASP, 121, 76

\bibitem[\protect\citeauthoryear{Case \&  Bhattacharya}{1998}]{cas98}	
Case G.L., Bhattacharya D., 1998, ApJ, 504, 761

\bibitem[\protect\citeauthoryear{Condon, Griffith \& Wright}{1993}]{cgw93}
Condon J.J., Griffith M.R., Wright A.L., 1993, AJ, 106, 1095

\bibitem[\protect\citeauthoryear{Condon et al.,}{1998}]{Con98}
Condon J.J., Cotton W.D., Greisen E.W., Yin Q.F., Perley R.A., Taylor G.B., Broderick J.J., 1998, AJ, 115, 1693.

\bibitem[\protect\citeauthoryear{Dopita et al.,}{2007}]{dop07}
Dopita M., Hart J.,  McGregor P., Oates P.,  Bloxham G.,  Jones D., 2007, ApSS, 310, 255

\bibitem[\protect\citeauthoryear{Drew et al.,}{2014}]{drew14}
Drew, J. E. et al., 2014, MNRAS, 440, 2036

\bibitem[\protect\citeauthoryear{Fesen et al.,}{1985}]{Fesen85}
Fesen R.A., Blair W.P., Kirshner R.P., 1985, ApJ, 292, 29

\bibitem[\protect\citeauthoryear{Fesen \& Milisavljevi\'c}{2010}]{Fes10}
Fesen R.A., Milisavljevi\'c D., 2010, AJ, 140, 1163

\bibitem[\protect\citeauthoryear{Frew \& Parker}{2010}]{Frew10}
Frew D.J., Parker Q. A., 2010, Publ.Astron.Soc.Aus, 27, 129

\bibitem[\protect\citeauthoryear{Frew et al.,}{2014a}]{Frew14}
Frew D.J., Boji\v{c}i\'c, I.S., Parker Q. A., Pierce M.J., Gunawardhana M.L.P., Reid W.A., 2014a, MNRAS, 440, 1080

\bibitem[\protect\citeauthoryear{Frew et al.,}{2014b}]{Frew14b}
Frew D.J., Boji\v{c}i\'c, I.S., Parker Q. A., Stupar, M., Wachter  S.,  DePew  K.,  Danehkar  A., Fitzgerald  M. T., Douchin, D.,  2014b, MNRAS, 440, 1345

\bibitem[\protect\citeauthoryear{Gaustad et al.,} {2001}]{gaus01}
Gaustad J.E., McCullough P.R., Rosing W., Van Buren D., 2001,
PASP, 113, 1326

\bibitem[\protect\citeauthoryear{Green}{2014}]{gree14}
Green D.A., 2014, BASI, 42, 47 (also available online at  http://www.mrao.cam.ac.uk/surveys/snrs/)

\bibitem[\protect\citeauthoryear{Grimani}{2004}]{grrim04}
 Grimani C., 2004, A\&A 418, 649

\bibitem[\protect\citeauthoryear{Helfand  et al.,}{2006}]{helf06}
Helfand D.J., Becker R.H., White R.L., Fallon A., Tuttle S., 2006, AJ, 131, 2525

\bibitem[\protect\citeauthoryear{McClure-Griffiths et al.,}{2005}]{mcC05}
McClure-Griffiths N. M., Dickey J. M., Gaensler B. M., Green A. J., Haverkorn
M.,  Strasser S. 2005, ApJS, 158, 178

\bibitem[\protect\citeauthoryear{Parker et al.,}{2005}]{parker05}
Parker Q.A., Phillipps S., Pierce M.J., Hartley M., Hambly N.C.,
Read M.A. et al., 2005, MNRAS, 362, 689

\bibitem[\protect\citeauthoryear{Parker et al.,}{2006}]{parker06}
Parker, Q. A., Acker, A., Frew, D.J., Hartley, M., Peyaud, A.E.J., Ochsenbein, F. et al., 2006, MNRAS, 373, 79

\bibitem[\protect\citeauthoryear{Parker et al.,}{2016}]{parker16}
Parker, Q. A., Boji\v{c}i\'c, I.S., Frew, D.J., 2016, Journal of Physics Conf.Ser., 728, 2008

\bibitem[\protect\citeauthoryear{Price et al.,}{2001}]{price01}
Price S.D., Egan M.P., Carey S.J., Mizuno D.R., Kuchar T.A., 2001, AJ, 121, 2819

\bibitem[\protect\citeauthoryear{Raymond}{1979}]{ray79}
Raymond J.C., 1979, ApJS, 39, 1

\bibitem[\protect\citeauthoryear{Reich, Reich \& F\"{u}rst,}{1990}]{reich90}
Reich W., Reich P., F\"{u}rst, E. 1990, A\&AS, 83, 539

\bibitem[\protect\citeauthoryear{Sabbadin, Minello \&  Bianchini}{1977}]{sabb77}
Sabbadin F., Minello S.  Bianchini A., 1977, A\&A, 60, 147

\bibitem[\protect\citeauthoryear{Sabin et al.,}{2013}]{sab13}
Sabin L.  et. al.,  2013, MNRAS, 431, 279

\bibitem[\protect\citeauthoryear{Stupar \& Parker}{2009}]{stu09}
Stupar M.,  Parker Q.A., 2009, MNRAS, 394, 1791

\bibitem[\protect\citeauthoryear{Stupar \& Parker}{2011}]{stu11}
Stupar M.,  Parker Q.A., 2011, MNRAS, 414, 2282

\bibitem[\protect\citeauthoryear{Stupar et al.,}{2005}]{stu05}
Stupar M., Filipovi\'c M. D., Jones P. A., Parker Q. A., 2005, AdSpR, 35,
1047

\bibitem[\protect\citeauthoryear{Stupar et al.,}{2007}]{stu07}
Stupar M., Parker Q.A., Filipovi\'c M.D., Frew D.J., Boji\v{c}i\'c I., Aschenbach
B., 2007, MNRAS, 381, 377

\bibitem[\protect\citeauthoryear{Stupar, Parker \& Filipovi\'c}{2008}]{stu08}
Stupar M.,  Parker Q.A.,  Filipovi\'c M.D., 2008, MNRAS, 390, 1037

\bibitem[\protect\citeauthoryear{Stupar, Parker \& Filipovi\'c}{201}]{stu11b}
Stupar M.,  Parker Q.A.,  Filipovi\'c M.D., 2011, ApSS, 332, 241

\bibitem[\protect\citeauthoryear{Voges et al.,}{2000}]{vog2000}
Voges W. et al., 2000, IAU Circular 7432

\bibitem[\protect\citeauthoryear{Wright et al.,}{2010}]{wri10}
Wright E.L. et al., 2010, AJ, 140, 1868

\end{thebibliography}
\end{document}